\documentclass{PoS}

\usepackage{amsmath}

\title{Electromagnetic splittings of hadrons from improved staggered
quarks in full QCD }

\ShortTitle{Electromagnetic splittings}

\author{\speaker{S. Basak},$^a$\thanks{Presently at NISER,
Bhubaneswar, Orissa 751005, India} ~A.~Bazavov, $^b$ C.~Bernard,
$^c$ C.~DeTar,$^d$ W.~Freeman,$^b$ Steven Gottlieb,$^a$ U.M.~Heller,$^e$
J.E.~Hetrick,$^f$ J.~Laiho,$^c$ L.~Levkova,$^d$ J.~Osborn,$^g$
R.~Sugar,$^h$ and D.~Toussaint$^b$ ~~(MILC Collaboration) \\
        \llap{$a$} Department of Physics, Indiana University\\
         Bloomington, IN 47405, USA\\
\llap{$^b$}
        Physics Department, University of Arizona\\ Tucson, AZ 85721,
USA\\
\llap{$^c$}
        Physics Department, Washington University\\
St. Louis, MO 63130, USA\\
\llap{$^d$}
        Physics Department, University of Utah\\
Salt Lake City, UT 84112, USA\\
\llap{$^e$}
        American Physical Society\\
 One Research Road, Box 9000, Ridge, NY 11961-9000, USA\\
\llap{$^f$}
        Physics Department, University of the Pacific\\
 Stockton, CA 95211, USA\\
\llap{$^g$}
        Argonne Leadership Computing Facility, Argonne National Laboratory\\
        Argonne, IL 60439, USA\\
\llap{$^h$}
        Physics Department, University of California\\
 Santa Barbara, CA 93106, USA\\
        E-mail: \email{sbasak@indiana.edu}
}

\abstract{
We present our initial study of the electromagnetic splittings
of charged and neutral mesons, and the violation of Dashen's
theorem. Hadron masses are calculated on MILC $N_f=2+1$ QCD
ensembles at lattice spacing $\approx 0.15$ fm, together with
quenched non-compact $U(1)$ configurations. The $\mathcal{O}(a^2)$
tadpole improved staggered quark (asqtad) action is used both
for the sea quarks and for six different valence quark masses.
Chiral extrapolations are performed using partially quenched
chiral perturbation theory incorporating electromagnetic
corrections.
}

\FullConference{The XXVI International Symposium on Lattice Field Theory \\
                 July 14 - 19, 2008\\
                 Williamsburg, Virginia, USA}

\begin{document}


\section{Introduction}
Lattice QCD has long been used to determine nonperturbative
hadronic properties; 
the spectrum, in particular, has been extensively investigated.
However, electromagnetic and isospin-breaking effects are usually
not included in those investigations. Although the strong interaction
is blind to the electromagnetic charges, we do in fact have mass
splittings of mesons and baryons in the real world which depend on
both isospin-breaking and electromagnetic interactions. Further,
it has been pointed out that the current evaluation of the light
quark masses, particularly the ratio $m_u/m_d$, suffers significant
uncertainties coming from the electromagnetic contributions to the
masses of $\pi$ and especially $K$ mesons \cite{milc_04}.

It is therefore important to include electromagnetic effects in the
lattice QCD simulations for more realistic spectrum calculations. The
pioneering attempt to calculate charged and neutral pion splittings
and the light quark masses from lattice QCD with electromagnetic
interactions was by Duncan, Eichten and Thacker \cite{duncan_em96}.
The photon fields were introduced in non-compact form and were treated
in the quenched approximation. The pseudoscalar meson masses were
calculated with the Wilson action and at different values of electric
charge. Recently, Blum {\em et al.} \cite{blum_emdw} have
calculated electromagnetic contributions to $\pi$ and $K$ mesons
mass splittings and determined the light quark masses with domain
wall quarks and $N_f=2$ QCD configurations at the physical value
of electric charge. They have also estimated the correction to
the Dashen's theorem \cite{dash_em} at $\mathcal{O}(\alpha m_q)$,
and found $\Delta_{\rm em}=0.337(40)$ or 0.264(43) depending on
their fitting range. Electromagnetic splittings of $\pi$ and $\rho$
mesons have also been calculated using a RG-improved gauge action
and a mean-field improved clover quark action at two different volumes
and three lattice spacings in Ref. \cite{namekawa_em}. Electromagnetic
effects on baryons have been discussed in Refs. \cite{duncan_embar,
blum_embar}; and calculation with unquenched photon fields in Ref.
\cite{duncan_emdyn}.

In this work, we study the electromagnetic mass splittings of
pseudoscalar mesons in the presence of a quenched electromagnetic
field and calculate the correction to the Dashen's theorem,
parametrized following the chiral perturbation
theory calculation of Urech \cite{urech_em}. The lattice data
is fitted using partially quenched chiral perturbation theory
formulas at $\mathcal{O}(p^4,\,e^2 m_q)$ given by Bijnens {\em
et al.} \cite{bij_pqcpt}. We have used a non-compact action for
the photon fields following \cite{duncan_em96}. The valence
quarks are asqtad \cite{milc_ksact} staggered quarks, and the
configurations are the MILC $N_f=2+1$ QCD configurations at
lattice spacing $0.15\,$ fm, with varying sea quark masses
\cite{milc_lat}. 


\section{Lattice QCD with electromagnetic interaction}
The mass differences among the members of hadron isomultiplets
arise from two sources: ({\em a}) strong isospin-breaking
contributions due to the difference in masses of the light
quarks and ({\em b}) different electromagnetic charges of the
quarks. Apart from a small isospin-breaking contribution of
order $(m_d-m_u)^2$, the $\pi^\pm-\pi^0$ mass splitting is
almost purely electromagnetic in origin; whereas it is the
isospin-breaking contributions that dominate for $K^0-K^\pm$.
Dashen's theorem \cite{dash_em} summarizes the electromagnetic
effects on meson masses: in the chiral limit to $\mathcal{O}(e^2)$,
electromagnetic interactions modify the leading order $M_{\pi^\pm}^2$
and $M_{K^\pm}^2$ masses while the $\pi^0$ and $K^0$ masses
remain unaffected,
\begin{equation}\begin{split}
M_{\pi^\pm}^2 = 2\hat{m}B_0 + A^{(1)}e^2(q_u-q_d)^2, & \hspace{0.4in}
M_{K^\pm}^2 = (\hat{m}+m_s)B_0 + A^{(1)}e^2(q_u-q_s)^2 \\
\left( M_{\pi^\pm}^2 - M_{\pi^0}^2 \right)_{\rm em} & \,=\,
\left( M_{K^\pm}^2 - M_{K^0}^2 \right)_{\rm em},
\end{split}\end{equation}
where $q_u$, $q_d$, and $q_s$ are the $u$, $d$, and $s$ quark
charges in units of $e$. At $\mathcal{O}(e^2m)$, however,
$M_{\pi^{\pm,0}}^2$ and $M_{K^{\pm,0}}^2$ can receive large and
different contributions, and the correction to Dashen's theorem
can parametrized as,
\begin{equation}
\Delta M_D^2 = \Delta M_K^2 - \Delta M_\pi^2 = \left( M_{K^\pm}^2
- M_{K^0}^2 \right)_{\rm em} - \left( M_{\pi^\pm}^2 - M_{\pi^0}^2
\right)_{\rm em},
\end{equation}
or as $\Delta_E$, defined by $\Delta M_K^2=(1+\Delta_E)\,\Delta
M_\pi^2$.
The partially quenched chiral perturbation theory relevant for QCD
+ QED with 2+1 dynamical flavors at NLO has recently been worked
out by Bijnens {\em et al.} \cite{bij_pqcpt}; it can be used to
perform fits to the lattice $M_{\pi,\, K}^2$ data in order to
determine the electromagnetic low energy constants. The pure
electromagnetic correction, relevant for calculating the correction
to Dashen's theorem, can be computed from the expression,
\begin{eqnarray}
\Delta M^2 \vert_{\rm em} &=& M_{\rm ps}^2(\chi_1,\chi_3,q_1,q_3)
- M_{\rm ps}^2(\chi_1,\chi_3,q_3,q_3) \nonumber \\
&-&  M_{\rm ps}^2(\chi_1,\chi_1,q_1,q_3) +
 M_{\rm ps}^2(\chi_1,\chi_1,q_3,q_3). \label{dash_eq}
\end{eqnarray}
In this notation \cite{bij_pqcpt}, the normalized quark mass is
$\chi_i = 2 B_0 m_i$, where $B_0$ is related to the quark-antiquark
vacuum expectation value in the chiral limit, and $ M_{\rm ps}^2(
\chi_1,\chi_3,q_1,q_3)$ denotes the squared pseudoscalar meson mass
having valence quark masses $\chi_1,\chi_3$ and valence charges
$q_1,q_3$ in units of electron charge $e$. In the isospin limit
$m_u=m_d$, when $\chi_1=m_u,\, \chi_3=m_s$ and the quark charges
$q_1=q_u,\,q_3=q_s$, we have $\Delta M_D^2 = \Delta M^2 \vert_{\rm
em}$. We fit the lattice data for $\Delta M_D^2$ according to
\begin{eqnarray}
\Delta M_D^2 &=& \mathcal{A}_1 (\chi_{13}-\chi_{11}) +
\mathcal{A}_2 \left[ \chi_{13}\log \left( \frac{\chi_{13}}{\mu^2}
\right) - \chi_{11} \log \left(\frac{\chi_{11}}{\mu^2} \right)
\right] \nonumber \\
&+& \mathcal{A}_3 \left[ \chi_{1f}\log\left(\frac{\chi_{1f}}{\mu^2}
\right) - \chi_{3f}\log\left(\frac{\chi_{3f}}{\mu^2}\right)
\right] \nonumber \\
&+& \mathcal{A}_4 \left[ \chi_{13} \left\{ 1 - \frac{1}{2}\log\left(
\frac{\chi_{13}}{\mu^2}\right) \right\} - \chi_{11} \left\{ 1 - 
\frac{1}{2} \log\left(\frac{\chi_{11}}{\mu^2}\right) \right\} \right],
\label{dash_fit}
\end{eqnarray} 
where, $\chi_{ij} \equiv (\chi_i+\chi_j)/2$ and $\mathcal{A}_i$ are
the constants to be determined from the fit. For the scale $\mu$ we
have used 1000MeV. The sea-quark dependence of the log-terms
appears through $\chi_{if}$, where the sea-quark index $f$ is
summed over all the sea quarks. 
Terms with the sea-quark charges contributing to $\Delta M_D^2$
do not involve unknown low energy constants (LECs) at this order
and hence are computable without lattice simulation \cite{bij_pqcpt}.
In our expression we have included the parameters  $\mathcal{A}_{2,3,4}$,
which are not present in Ref. \cite{bij_pqcpt}, in front of the
logarithm terms. This is because we expect the continuum expression
to be modified  by discretization effects on the quite coarse
lattices we are using. Finite volume effects, due to the masslessness
of the photon, may also be rather significant.  At this stage,
therefore, we can obtain at best a rough estimate of the quantity
$\Delta M_D^2$; a precise determination will require lattice data
at finer lattice spacings and larger volumes.


\section{Numerical results}
Phenomenologically relevant quantities, such as the mass-squared
differences of charged and neutral mesons and the correction to Dashen's
theorem, can be calculated from partially quenched lattice QCD
where photons are treated in the quenched approximation. Therefore,
we can use the existing $SU(3)$ configurations generated without
dynamical photons. For unquenched QCD gluon configurations, we
have used four ensembles of existing $a\approx 0.15\,$fm MILC lattices
with 2+1 dynamical flavors, the details of which are provided in
Table 1. The lattice spacing is set by the static quark potential
\cite{milc_lat}.
\begin{table}[h]
\begin{center}
\begin{tabular}{|c|c|c|c|c|} \hline
Size & $\beta=10/g^2$ & $a\hat{m}^\prime/am_s^\prime$ & $a$ (fm)
& \# Cfgs \\ \hline
$20^3\times 48$ & 6.566 & 0.00484/0.0484 & $\approx$ 0.15 & 200 \\
$16^3\times 48$ & 6.572 & 0.0097/0.0484 & $\approx$ 0.15 & 400 \\
$16^3\times 48$ & 6.586 & 0.0194/0.0484 & $\approx$ 0.15 & 400 \\
$16^3\times 48$ & 6.600 & 0.0290/0.0484 & $\approx$ 0.15 & 400 \\ \hline
\end{tabular}
\end{center}
\caption{
$N_f=2+1$ MILC lattices used in this project.}
\end{table}

\noindent
The quenched photon configurations $\{A_\mu(n)\}$ are generated
from the non-compact $U(1)$ action,
\begin{equation}
S_{\rm em}=\frac{1}{4} \sum_{n\mu\nu} \left( \partial^r_\mu
A_\nu(n) - \partial^r_\nu A_\mu(n) \right)^2,
\end{equation}
subjected to Coulomb gauge fixing $\partial^l_i A_i(n)=0$.
Additional global gauge fixing is done to ensure that Gauss's law
is satisfied. In
momentum space, the action is Gaussian distributed, and the
scalar and vector potentials $[A_0(p)$, $\vec{A}(p)]$ can be
generated independently of each other from Gaussian distributed
random numbers. The coordinate-space Coulomb-gauge photon
configurations are recovered by FFT.
The necessary valence quark propagators are calculated with wall
sources in these $SU(3) \times U(1)$ background fields. We
calculate the pseudoscalar meson propagator with nine different
valence quark masses, $0.1m_s^\prime \leq m_q \leq 1.0m_s^\prime$.
All the fits we report here are obtained by correlated $\chi^2$-fit;
the errors are obtained from a jackknife analysis. The meson masses are
extracted from the exponential fall-off of the meson propagators in
the time range 9 -- 24, taking into account the correlations
among the time slices. We have ignored the contribution from
the disconnected diagram that effects the neutral
pion mass.

\begin{figure}
\begin{center}
\vskip 0.1in
\includegraphics[width=0.54\textwidth]{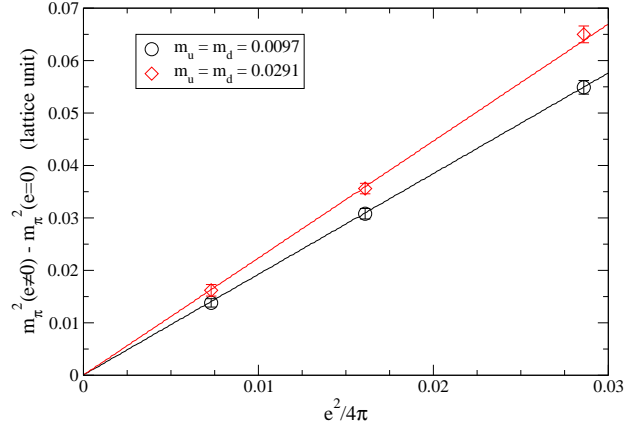}
\end{center}
\vskip -0.25in
\caption{The $e^2$ dependence of mass-squared electromagnetic splittings
of the pion.}
\label{pi2_alpha}
\end{figure}

\begin{figure}
\begin{center}
\vskip 0.22in
\includegraphics[width=0.54\textwidth]{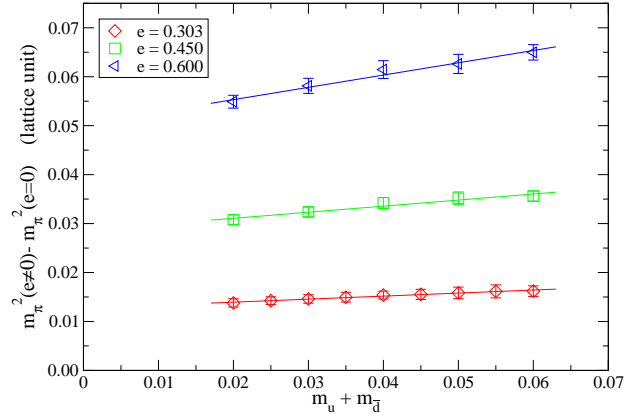}
\end{center}
\vskip -0.25in
\caption{The $\mathcal{O}(e^2m)$ behavior of mass-squared
electromagnetic splittings of the pion.}
\label{pi2_mMe0}
\end{figure}

We first try to estimate the $\mathcal{O} (e^2)$ and $\mathcal{O}
(e^2m)$ contribution to pseudoscalar masses at three different
(physical and non-physical) values of electric charge,
\begin{equation}
M_\pi^2(e\ne 0) - M_\pi^2(e=0) = \mathcal{A}_0 e^2 (q_u-q_d)^2 +
\mathcal{O}(e^2m). \label{pi2_alpha_split}
\end{equation}
In Fig.~\ref{pi2_alpha}, we plot the dependence of $\pi$ mass-squared 
splittings corresponding to Eq. (\ref{pi2_alpha_split}). A straight
line fit describes the $\mathcal{O}(e^2)$ behavior fairly well over
the full range of electric charges that are examined. In a further
test to ascertain the $\mathcal{O} (e^2m)$ or higher order effects,
in Fig. \ref{pi2_mMe0}, we plot the variation of the same pion
mass-squared splittings against quark masses. The slope of each of
line, corresponding to different electric charges, nicely matches
to its $e^2$ value.

\begin{figure}[ht]
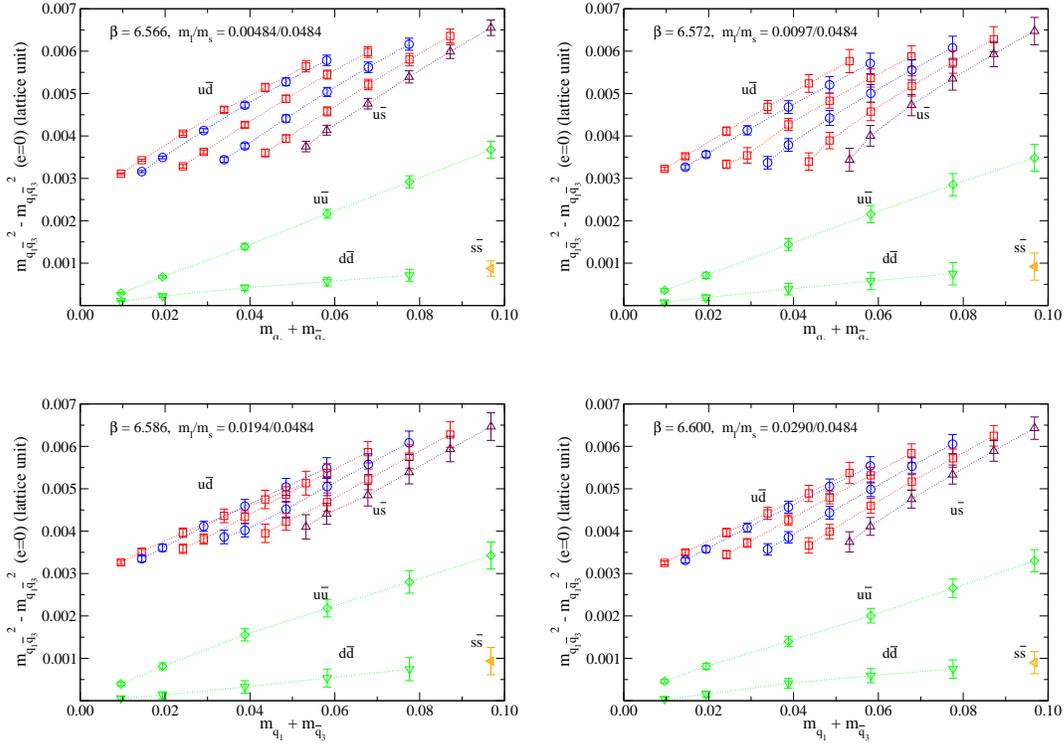

\begin{tabular}{ll}
\includegraphics[width=0.45\textwidth]{b6566_pi2_e1dme0.eps} &
\includegraphics[width=0.45\textwidth]{b6572_pi2_e1dme0.eps} \\
&\\
\includegraphics[width=0.45\textwidth]{b6586_pi2_e1dme0.eps} &
\includegraphics[width=0.45\textwidth]{b6600_pi2_e1dme0.eps} 
\end{tabular}
\caption{The mass-squared splittings between pseudoscalar mesons with
physical value of electric charge $e^2=4\pi\alpha_{\rm em}$ and $e^2=0$.
Each panel corresponds to different $\beta$ and light sea-quark mass
$am_l^\prime$ for fixed sea-strange quark mass $am_s^\prime=0.0484$.
The labels $q_1\bar{q}_2$ denote the quark charges. We write $\bar s$
instead  of ~$\bar d$ whenever $m_{q_2}$ is equal to $m'_s$, the strange
sea-quark mass in the simulation.}
\label{pi2_e1dme0}
\end{figure}

In a series of four plots in Fig. \ref{pi2_e1dme0}, we show the
meson mass splittings between mesons computed with physical electric
charge $e^2=4\pi\alpha_{\rm em}$ and with $e^2=0$. The labels $q_1
\bar{q}_2$ in the plots denote quark charges, {\em i.e.}, $u\bar{d}$
indicates quarks with charge $q_u=2e/3$ and $q_d=-1e/3$. The points
using the same symbol and color are obtained by varying the $q_2$
quark mass. The lines are merely to guide the eye. We have not yet
fit the squared meson masses to the Bijnens form based on partially
quenched chiral perturbation theory (and hence not yet extracted
the electromagnetic LECs) \cite{bij_pqcpt}. These LECs will be
important in determining the physical isomultiplet splittings in
mesons and therefore in calculating quark masses.

\begin{figure}[h]
\begin{center}
\vskip 0.12in
\includegraphics[width=0.54\textwidth]{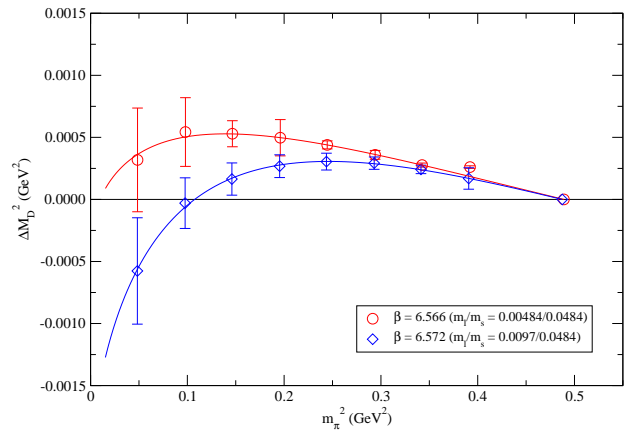}
\end{center}
\caption{Correction to Dashen's theorem, the difference of
electromagnetic contributions between $K$ and $\pi$ masses,
as a function of the LO $\pi$ mass squared (equivalent 
to the $\pi$ mass squared with $e^2=0$).}
\label{dash_ps}
\end{figure}

The electromagnetic LECs can be used to obtain the correction
to Dashen's theorem, but we have estimated this correction
directly by chiral extrapolation of Eq.~(\ref{dash_eq}) using
Eq.~(\ref{dash_fit}). We put $q_1=q_u,\, q_3=q_s$ in Eq.
(\ref{dash_eq})  and extrapolate/interpolate to $\chi_1=2B_0m_u,
\, \chi_3=2B_0m_s$. Figure \ref{dash_ps} is our preliminary result
for the violation of Dashen's theorem, where we have plotted
the $\Delta M_D^2$ for two ensembles, $\beta=6.566$ and 6.572,
as a function of $M_\pi^2$, obtained with $e^2=0$, in physical
units. The data from the remaining two QCD ensembles at heavier
quark mass are still being analyzed. We have determined the
coefficients $\mathcal{A}_i$ in Eq. (\ref{dash_fit}) using our
lattice data for pseudoscalar masses $\chi_{ij}$, after which
chiral extrapolation is performed at the experimental masses
for $\pi$ and $K$.  Our preliminary result for the deviation
from Dashen's theorem is $7.0\times 10^{-4} < \Delta M_D^2\,/
({\rm GeV}^2) < 1.8 \times 10^{-3}$. This may be compared with
the estimate in Ref \cite{bij_pqcpt}:  $1.07 \times 10^{-3}$.
Although we find that our
value is consistent with the phenomenological estimate, the error
in our result is large and does not yet include the systematic
errors due to discretization and finite volume effects. We hope
to reduce the error by increasing the statistics and extending
the calculation to finer lattices and larger volumes. Fitting the
meson splittings  in Fig.~\ref{pi2_e1dme0}  to determine the
electromagnetic LECs directly may also provide better control
over our errors.

\section{Conclusions and outlook}
In this work we have calculated the electromagnetic mass splittings
of pseudoscalar masses with $\mathcal{O}(a^2)$ improved staggered
fermions on $a\approx0.015\,$fm $N_f=2+1$ MILC $SU(3)$ ensembles
and Coulomb-gauge $U(1)$ configurations. We have determined the
correction to Dashen's theorem at the physical point (in the isospin
limit), but with relatively large errors, {\em i.e.}, a wide range
for $\Delta M_D^2$. Our determination of meson and baryon isomultiplet
splittings and quark masses is in progress.
Our immediate goal is using finer lattices to reduce the error in
the correction to Dashen's theorem enough to bring
down the present error in $m_u/m_d$.

\end{document}